\documentclass{revtex4}
\usepackage{graphicx}
\usepackage{fancyhdr}
\usepackage{amsmath}
\newcommand{\SU}{{\text{SU}}}
\newcommand{\eV}{{\text{eV}}}

\newcommand{\diag}{{\text{diag}}}
\newcommand{\BR}{{\text{BR}}}
\newcommand{\SM}{{\text{SM}}}
\newcommand{\MNS}{{\text{MNS}}}

\usepackage{color,multirow}

\begin{document}

\title{Neutrino Mass in TeV-Scale New Physics Models}

%

\author{Hiroaki Sugiyama%
~\footnote{
 Current affiliation of the author is
Maskawa Institute for Science and Culture,
Kyoto Sangyo University,
Kyoto 603-8555, Japan.
}}
\affiliation{Department of Physics, University of Toyama,
Toyama 930-8555, Japan}

\begin{abstract}
 This is a short review about relations
between new scalars and
mechanisms to generate neutrino masses.
 We investigate leptohilic scalars
whose Yukawa interactions are only with leptons.
 We discuss possibilities that
measurements of their leptonic decays
provide information on
how neutrino masses are generated
and on parameters in the neutrino mass matrix
(e.g.\ the lightest neutrino mass).
\end{abstract}

\maketitle

\thispagestyle{fancy}



\section{Introduction}
 In the standard model of particle physics~(SM),
neutrinos are regarded as massless particles.
 However,
existence of nonzero masses of neutrinos
has been established by a sequence of
success of neutrino oscillation measurements%
~\cite{Ref:solar-v, Ref:atom-v, Ref:acc-v, Ref:acc-app-v,
Ref:short-reac-v, Ref:long-reac-v}.
 If neutrino masses are generated similarly to the other fermion masses
via a Yukawa interaction (introducing right-handed neutrinos $\nu_R^{}$)
with the $\SU(2)_L$-doublet scalar field $\Phi_\SM$ in the SM,
the Yukawa coupling constants must be extremely small ($\sim 10^{-12}$).
 Since the coupling constant
seems too different from the other Yukawa coupling constants
to regard it as natural,
we might expect that neutrino masses are generated
in a different mechanism.
 The most famous example will be 
the seesaw mechanism~\cite{Ref:Type-I}
where extremely heavy gauge-singlet fermions are introduced.

 In this short review,
we consider extensions of the SM with new scalars
(in contrast with the seesaw models)
along the viewpoint "Higgs as a probe of new physics" of this workshop.
 Along my viewpoint "neutrino as a guide to new physics",
we investigate only leptophilic scalars
which couple only with leptons among fermions
because such scalars would contribute to
mechanism of generating neutrino masses.
 In order to give predictions,
it will not be preferred
that two new particles appear in an interaction
(Lepton--"New fermion"--"New scalar").
 Therefore,
let us concentrate on the following types
of Yukawa interactions:
\begin{center}
 Lepton--Lepton--"Leptophilic scalar" .
\end{center}
 The leptophilic scalars here are assumed to be light enough (TeV-scale)
to be produced at collider experiments.
 If the Yukawa interaction relates to the neutrino mass matrix,
the flavor structure of decays of the new scalar into leptons
would be predicted by using
current knowledge on the neutrino oscillation parameters.
 If the prediction is experimentally confirmed in the future,
we would obtain information on the mechanism of the neutrino mass generation
and on parameters (e.g.\ the lightest neutrino mass)
which cannot be measured in neutrino oscillation experiments.

 We deal with $\SU(2)_L$-singlet, doublet, and triplet scalar fields.
 Their Yukawa interactions are listed in Table~\ref{Tab:table}.
 For simplicity, mixings between scalars are ignored
throughout this article.
 Leptophilic neutral scalars in doublet and triplet fields
are not discussed in this article
because their decays via Yukawa interactions are into neutrinos
which do not provide information on the flavor structure.

\renewcommand\baselinestretch{2}
\begin{table}[t]
\begin{tabular}{l||l||l|l}
{}
 & \multirow{2}{*}{\hspace*{13mm} Yukawa interaction}
 & \multicolumn{2}{c}{Decay into leptons}
\\\cline{3-4}
{}
 &
 & \ Singly charged \
 & \ Doubly charged \
\\\hline\hline
$\SU(2)_L$-singlet \
 & \
   $
    f_{\ell\ell^\prime}^{}
    \Bigl[
     \overline{L_\ell^c}\,i\sigma_2\, L_{\ell^\prime}^{}\, s^+
    \Bigr]
   $ , \quad
   $
    f_{\ell\ell^\prime}^\prime
    \Bigl[ \overline{(\ell_R)^c}\, \ell_R^\prime\, s^{++} \Bigr]
   $ \
 & \ $s^+ \to \overline{\ell_L^{}}\, \overline{\nu_{\ell^\prime L}^{}}$ \
 & \ $s^{++} \to \overline{\ell_R^{}}\, \overline{\ell^\prime_R}$ \
\\[1mm]
\hline
$\SU(2)_L$-doublet \
 & \
  $
    y_{i\ell}^{}
    \Bigl[
     \overline{\nu_{iR}^{}}\, \Phi_\nu^T i\sigma_2\, L_\ell
    \Bigr]
   $ \
 & \ $\phi_\nu^+ \to \overline{\ell_L}\, \nu_{iR}^{}$ \
 & {}
\\[1mm]
\hline
$\SU(2)_L$-triplet \
 & \
  $
    h_{\ell\ell^\prime}^{}
    \Bigl[
     \overline{L_\ell^c}\, i\sigma_2\, \Delta\, L_{\ell^\prime}^{}
    \Bigr]
   $ \
 & \ $\Delta^+ \to \overline{\ell_L^{}}\, \overline{\nu_{\ell^\prime L}^{}}$ \
 & \ $\Delta^{++} \to \overline{\ell_L^{}}\, \overline{\ell^\prime_L}$ \
\end{tabular}
\caption{
 Yukawa interactions of leptophilic scalar fields
and leptonic decays of their singly-charged and doubly-charged components.
}
\label{Tab:table}
\end{table}
\renewcommand\baselinestretch{1}

\section{Basics}
\label{Sec:basics}

 Neutrinos $\nu_{\ell L}^{}~(\ell = e, \mu, \tau)$ in the flavor basis
are superpositions of mass eigenstates $\nu_{i L}^{}$:
$\nu_{\ell L}^{} = \sum_i (U_\MNS)_{\ell i}\, \nu_{i L}^{}$,
where the unitary matrix $U_\MNS$
is the so-called Maki-Nakagawa-Sakata matrix~\cite{Maki:1962mu}.
 When the neutrino mass term is
$(m_\nu^{})_{i\ell} \overline{\nu_{iR}^{}} \nu_{\ell L}^{}$,
neutrinos are referred to as the Dirac neutrinos.
 The mass matrix for Dirac neutrinos
is diagonalized with $U_\MNS$ as
$m_\nu^{} U_\MNS = \diag( m_1, m_2, m_3)$,
where mass eigenvalues $m_i$ are taken to be real and positive.
 On the other hand,
if the neutrino mass term is
$(m_\nu^{})_{\ell\ell^\prime}
\overline{(\nu_{\ell L}^{})^c} \nu_{\ell^\prime L}^{}$,
we call the neutrinos as the Majorana neutrinos
which break the lepton number conservation.
 The Majorana neutrino mass matrix
is diagonalized as
$U_\MNS^T m_\nu^{} U_\MNS
= \diag( m_1, m_2 e^{i\alpha_{21}^{}}, m_3 e^{i\alpha_{31}^{}})$,
where $\alpha_{21}$ and $\alpha_{31}$ are Majorana phases%
~\cite{Ref:M-phase-1, Schechter:1980gr}
which are physical parameters only for Majorana neutrinos.
 The matrix $U_\MNS$ can be parameterized as
\begin{eqnarray}
U_\MNS
=
\begin{pmatrix}
 1 & 0 & 0\\
 0 & c_{23} & s_{23}\\
 0 & -s_{23} & c_{23}
\end{pmatrix}
\begin{pmatrix}
 c_{13} & 0 & s_{13} e^{-i\delta}\\
 0 & 1 & 0\\
 -s_{13} e^{i\delta} & 0 & c_{13}
\end{pmatrix}
\begin{pmatrix}
 c_{12} & s_{12} & 0\\
 -s_{12} & c_{12} & 0\\
 0 & 0 & 1
\end{pmatrix} ,
\end{eqnarray}
where $s_{ij}^{}$ and $c_{ij}^{}$
stand for $\sin\theta_{ij}$ and $\cos\theta_{ij}$,
respectively.
 Current data of neutrino oscillation measurements%
~\cite{Ref:solar-v, Ref:atom-v, Ref:acc-v, Ref:acc-app-v,
Ref:short-reac-v, Ref:long-reac-v}
constrains mixing angles
and squared-mass differences ($\Delta m^2_{ij} \equiv m_i^2 - m_j^2$)
as
\begin{eqnarray}
\sin^2{2\theta_{23}} \simeq 1 , \quad
\sin^2{2\theta_{13}} \simeq 0.089 , \quad
\sin^2{2\theta_{12}} \simeq 0.85 ,\\
\Delta m^2_{21} \simeq 7.5\times 10^{-5}\,\eV^2 , \quad
|\Delta m^2_{31}| \simeq 2.3\times 10^{-3}\,\eV^2 .
\end{eqnarray}

\section{Singly Charged Scalar}
\label{Sec:singly}

\subsection{$\SU(2)_L$-singlet I}
\label{Subsec:singly-singlet-I}

 A singly charged scalar $s^+$
of an $\SU(2)_L$-singlet with the hypercharge $Y=1$
couples to the lepton doublet
$L_\ell = (\nu_{\ell L}^{} , \ell_L)^T$ as
\begin{eqnarray}
f_{\ell\ell^\prime}^{}
\Bigl[
 \overline{L_\ell^c}\,i\sigma_2\,
 L_{\ell^\prime}^{}\, s^+
\Bigr]
=
 - 2 f_{\ell\ell^\prime}^{}
 \Bigl[
  \overline{(\ell_L)^c}\, \nu_{\ell^\prime L}^{} s^+
 \Bigr] ,
\label{Eq:s-Yukawa}
\end{eqnarray}
where the matrix of Yukawa coupling constants
is antisymmetric~($f = -f^T$)
and $\sigma_i (i=1\text{-}3)$ are the Pauli matrices.
 The scalar $s^+$ is introduced in e.g.\
the so-called Zee model~\cite{Ref:Zee}
where light Majorana neutrino masses
are generated at the one-loop level.
 The simplest version of the Zee model~\cite{Ref:Zee, Wolfenstein:1980sy}
where there is no Flavor-Changing-Neutral-Current~(FCNC)
was excluded by neutrino oscillation measurements%
~(See e.g.\ Ref.~\cite{He:2003ih}).

 Motivated by the original version of the Zee model where the FCNC exists,
we consider
a Majorana neutrino mass matrix $(m_\nu^{})_{\ell\ell^\prime}^{}$
of the following structure in the flavor basis:
\begin{eqnarray}
(m_\nu^{})_{\ell\ell^\prime}^{}
=
 \Bigl(
  X m_\ell^\diag f
  + (X m_\ell^\diag f)^T
 \Bigr)_{\ell\ell^\prime} ,
\label{Eq:mnu-ZM-1}
\end{eqnarray}
where $X$ is an arbitrary matrix
and $m_\ell^\diag \equiv \diag( m_e^{} , m_\mu^{} , m_\tau^{} )$.
 Let us assume that contributions of $m_e^{}$ and $m_\mu^{}$
to $m_\nu^{}$ are negligible.
 Then
Eq.~\eqref{Eq:mnu-ZM-1} is simplified as
\begin{eqnarray}
(m_\nu^{})_{\ell\ell^\prime}^{}
\simeq
 m_\tau^{}
 \Bigl(
  X_{\ell \tau} f_{\tau \ell^\prime}
  + f_{\ell\tau} X_{\tau\ell^\prime}
 \Bigr) .
\label{Eq:mnu-ZM-2}
\end{eqnarray}
 Note that $m_\nu^{}$ is a rank-2 matrix under this assumption
although each term in the right-hand side of Eq.~\eqref{Eq:mnu-ZM-2}
is a rank-1 matrix.
 Thus,
the lightest neutrino becomes massless ($m_1=0$ or $m_3=0$)
while the other two have non-zero masses as required.
 The $m_\nu^{}$ includes four parameters:
$f_{e\tau} X_{e\tau}$ , $f_{\mu\tau}/f_{e\tau}$,
$X_{\mu\tau}/X_{e\tau}$, and $X_{\tau\tau}/X_{e\tau}$.
 The latter three combinations can be expressed with
neutrino mixing parameters by using conditions
that $U_\MNS$ diagonalizes the $m_\nu$
(three off-diagonal parts must be zero).
 Since $f_{e\tau} X_{e\tau}$ is an overall factor
for neutrino mass eigenstates,
the ratio of nonzero mass eigenstates
does not depend on $f_{e\tau} X_{e\tau}$.
 We see that $m_3 = 0$
(the so-called inverted hierarchy where $m_1/m_2 \simeq 1$)
is allowed while $m_1=0$
(the so-called normal hierarchy where $m_3/m_2 \gg 1$)
cannot be obtained.
 When we use simple values
$\sin^2{\theta_{23}} = 1/2$ and $\sin^2{\theta_{12}} = 1/3$
which are almost consistent with neutrino oscillation measurements,
$m_1/m_2 \simeq 1$ results in
\begin{eqnarray}
\sin^2{2\theta_{13}} \simeq 0.11 , \quad
\delta \simeq \pi , \quad
\alpha_{21}^{} \simeq \pi .
\label{Eq:param-Zee}
\end{eqnarray}
 The "predicted" value $\sin^2{2\theta_{13}} \simeq 0.11$
seems reasonably agree with observations%
~\cite{Ref:acc-app-v,Ref:short-reac-v}.
 See Ref.~\cite{He:2011hs} for more detailed analysis
of the Zee model.

 Partial decay widths $\Gamma_\ell^{(s)}$
for $s^+ \to \overline{\ell_L^{}}\, \overline{\nu_L^{}}$
(where neutrino species are summed)
are proportional to $\sum_{\ell^\prime} |f_{\ell\ell^\prime}|^2$.
 In the discussion above,
$f_{\mu\tau}/f_{e\tau}$ is constrained
by neutrino mixing parameters.
 Thus,
we obtain the following "prediction" for a combination
of branching ratios
$\BR_\ell^{(s)}
\equiv \BR(s^+ \to \overline{\ell_L^{}}\, \overline{\nu_L^{}})$:
\begin{eqnarray}
\frac{ \BR_\tau^{(s)} }{ \BR_e^{(s)} - \BR_\mu^{(s)} }
\simeq \frac{ 1 + 2 s_{13}^2 }{ 1 - 2 s_{13}^2 }
\simeq 1 .
\end{eqnarray}
 If this relation is confirmed experimentally,
neutrino mass matrix might be of Eq.~\eqref{Eq:mnu-ZM-2}
with parameters in Eq.~\eqref{Eq:param-Zee}.

\subsection{$\SU(2)_L$-singlet II}
\label{Subsec:singly-singlet-II}

 The singly-charged scalar $s^+$
is utilized also in the so-called Zee-Babu model~\cite{Ref:ZBM}
where Majorana neutrino masses are generated at the two-loop level.
 Another example is a model of loop-induced Dirac neutrino masses%
~\cite{Ref:loop-Dirac-1, Ref:loop-Dirac-2}.
 Neutrino mass matrix might be given by
\begin{eqnarray}
(m_\nu^{})_{\ell\ell^\prime}
=
 (f^T X_s f)_{\ell\ell^\prime}
\label{Eq:s-Majorana}
\end{eqnarray}
for Majorana neutrinos (similarly to the Zee-Babu model)
or
\begin{eqnarray}
(m_\nu^{})_{i\ell}
\Bigl[ \overline{\nu_{iR}^{}}\, \nu_{\ell L}^{} \Bigr]
=
 (X f)_{i\ell}
 \Bigl[ \overline{\nu_{iR}^{}}\, \nu_{\ell L}^{} \Bigr]
\label{Eq:s-Dirac}
\end{eqnarray}
for Dirac neutrinos
(similarly to the model in Refs.~\cite{Ref:loop-Dirac-1, Ref:loop-Dirac-2}).
 The matrix $X_s$ is symmetric while $X$ is arbitrary.
 Ratios of three elements of $f_{\ell\ell^\prime}$
can be easily obtained as functions of neutrino mixing parameters%
~\cite{Ref:ZBM-LHC-1}~(See also Refs.~\cite{Ref:ZBM-LHC-2, Ref:loop-Dirac-2}).
 Results are the same for Eqs.~\eqref{Eq:s-Majorana} and \eqref{Eq:s-Dirac}.
 For
$\sin^2{\theta_{23}} = 1/2$ and $\sin^2{\theta_{12}} = 1/3$,
we obtain
\begin{eqnarray}
\BR_e^{(s)} : \BR_\mu^{(s)} : \BR_\tau^{(s)}
=
 \begin{cases}
  2 : 5 : 5 & (\text{for $m_1 = 0$})\\
  2 : 1 : 1 & (\text{for $m_3 = 0$})
 \end{cases} .
\end{eqnarray}
 If experiments confirm these ratios,
the structure of the neutrino mass matrix
in Eq.~\eqref{Eq:s-Majorana} or \eqref{Eq:s-Dirac}
might be true.

\subsection{$\SU(2)_L$-doublet}
\label{Subsec:singly-doublet}

\begin{figure}
\begin{center}
\includegraphics[scale=0.33,angle = -90]{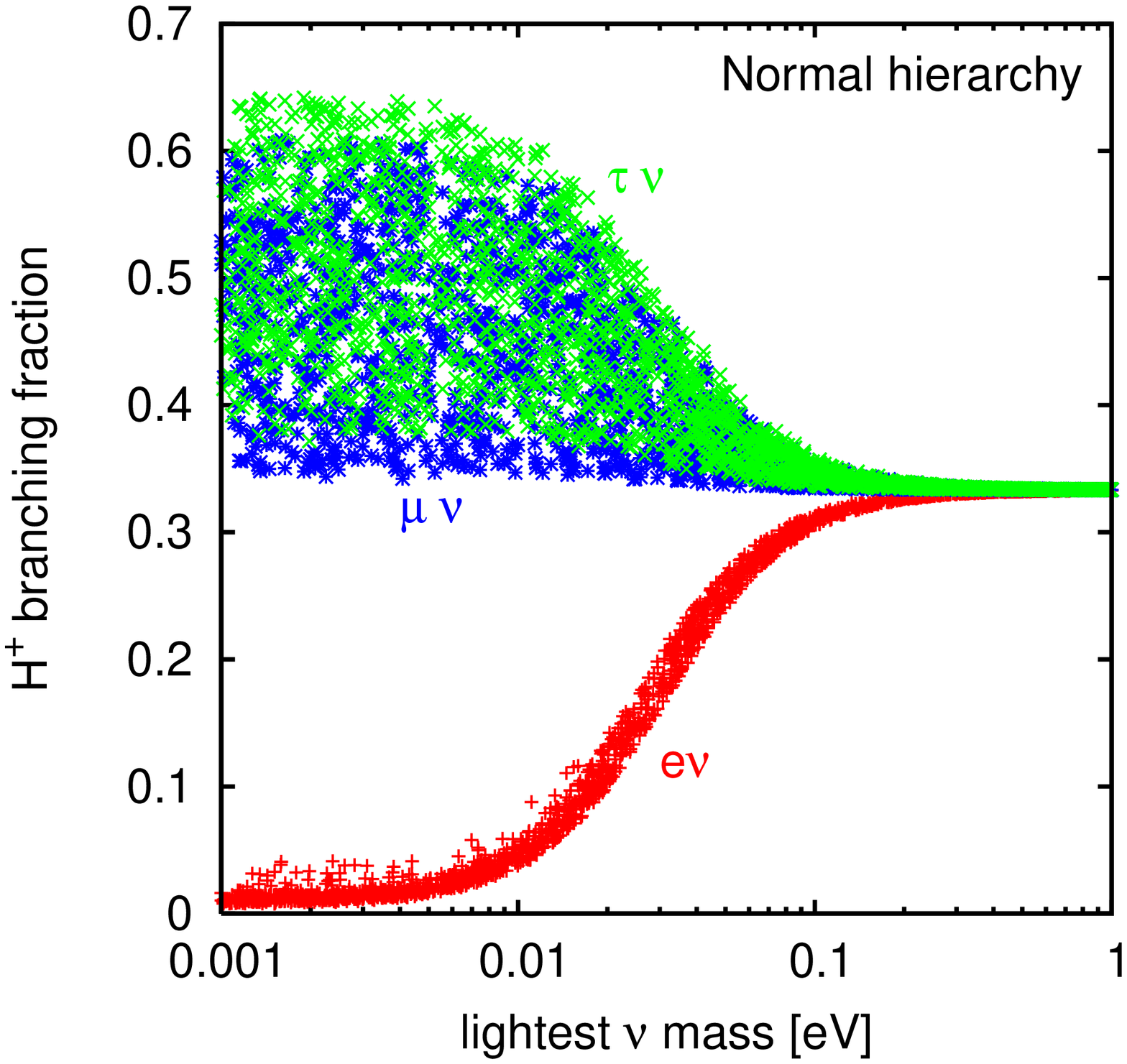}{} \ \ \
\includegraphics[scale=0.33,angle = -90]{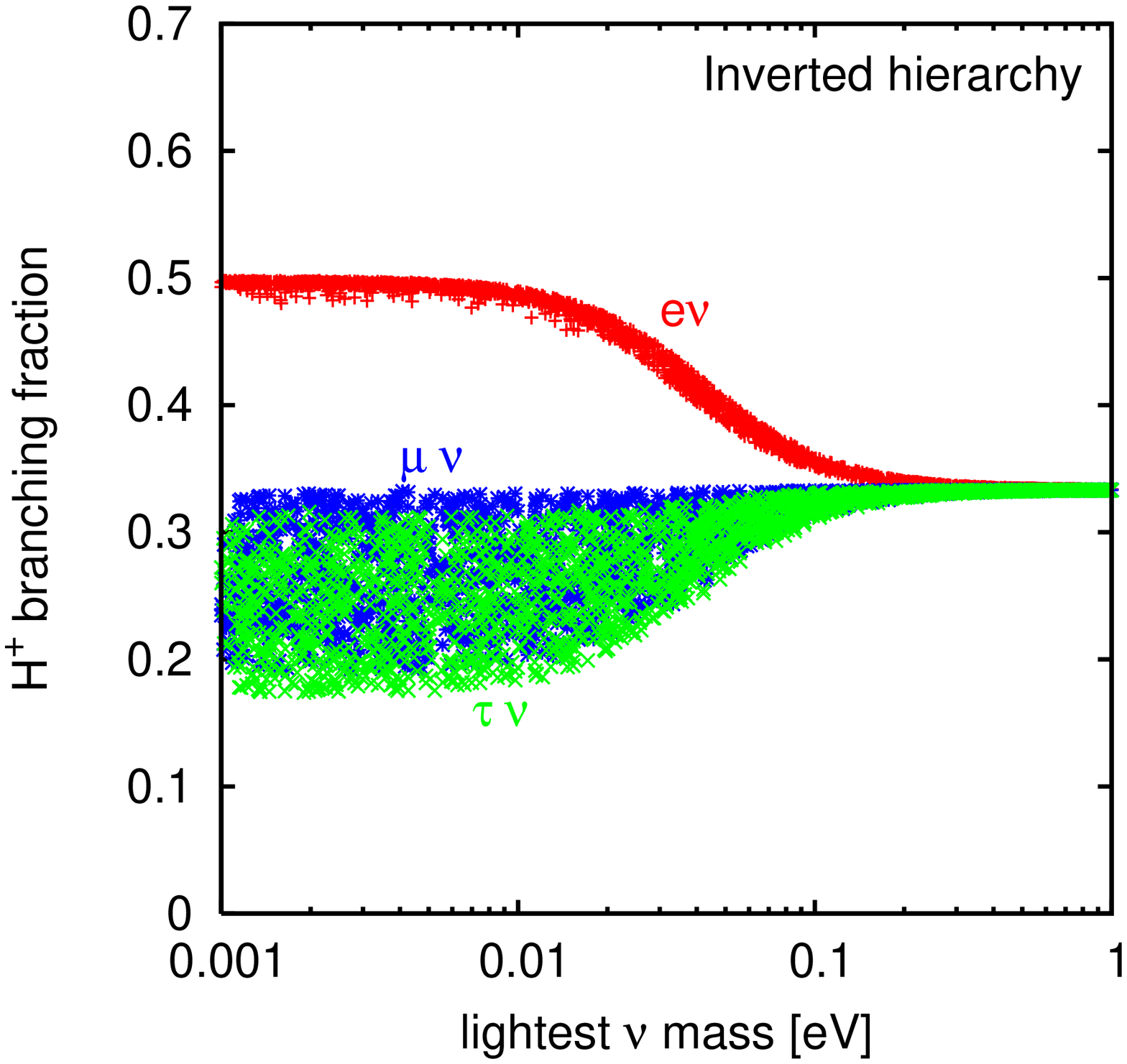}{}
\end{center}
\caption{
 Behaviors of $\BR(\phi_\nu^+ \to \overline{\ell_L^{}} \nu_R^{})$
with respect to $m_1$ (left panel) and $m_3$ (right panel).
 This figure is taken from Ref.~\cite{Ref:nuTHDM-2}.
}
\label{Fig:H+}
\end{figure}

 Neutrinos might obtain their Dirac masses
via a vacuum expectation value~(vev)
of an additional $\SU(2)_L$-doublet scalar field $\Phi_\nu$%
~\cite{Ref:nuTHDM-1, Ref:nuTHDM-2}.
 We refer to the model as the neutrinophilic two-Higgs-doublet model.
 The Yukawa interaction of neutrinos with $\Phi_\nu$ is written as
\begin{eqnarray}
- (y_\nu^{})_{i\ell}^{}
 \Bigl[
  \overline{\nu_{iR}^{}}\,
  \Phi_\nu^T i\sigma_2\, L_\ell
 \Bigr]
=
 (y_\nu^{})_{i\ell}^{}
 \Bigl[
  \overline{\nu_{iR}^{}}\, \nu_{\ell L}^{} \phi_\nu^0
 \Bigr]
 -
 (y_\nu^{})_{i\ell}^{}
 \Bigl[
  \overline{\nu_{iR}^{}}\, \ell_L \phi_\nu^+
 \Bigr] ,
\end{eqnarray}
where $\nu_{iR}^{}$ are right-handed components
of mass eigenstates $\nu_i$
(so, we do not regard $\nu_{iR}^{}$ as new particles here).
 The mass matrix of the Dirac neutrinos
is simply given by
$(m_\nu^{})_{i\ell} = \langle \phi_\nu^0 \rangle (y_\nu^{})_{i\ell}^{}$.
 The branching ratios $\BR(\phi_\nu^+ \to \overline{\ell_L^{}} \nu_R^{})$,
where neutrino species are summed,
are proportional to $(m_\nu^\dagger m_\nu^{})_{\ell\ell}^{}$.
 Figure~\ref{Fig:H+} (taken from Ref.~\cite{Ref:nuTHDM-2})
shows behaviors of these branching ratios
with respect to the lightest neutrino mass
($m_1$ for the left panel, and $m_3$ for the right one)
for the case $\phi_\nu^+$ decays only into leptons.
 By measuring $e\nu$ mode,
it would be possible to extract information
on the value of the lightest neutrino mass
and on whether $m_1 < m_3$~(left panel in Fig.~\ref{Fig:H+}) or not.
 If experiments show that
$\BR(H^+ \to \overline{\tau_L^{}} \nu_R^{})$
is very different from
$\BR(H^+ \to \overline{\mu_L^{}} \nu_R^{})$,
the charged scalar might not contribute
to the mechanism of generating neutrino masses
in a simple way.

\subsection{$\SU(2)_L$-triplet}
\label{Subsec:singly-triplet}

 A singly charged scalar exists in
an $\SU(2)_L$-triplet field $\Delta$ with $Y=1$,
which can be expressed as
\begin{eqnarray}
\Delta
\equiv
 \begin{pmatrix}
  \Delta^+/\sqrt{2}
  & \Delta^{++}\\
  \Delta^0
  & -\Delta^+/\sqrt{2}
 \end{pmatrix} .
\end{eqnarray}
 The triplet scalar field interacts with
the lepton doublet as
\begin{eqnarray}
h_{\ell\ell^\prime}^{}
\Bigl[
 \overline{L_\ell^c}\, i\sigma_2\,
 \Delta\, L_{\ell^\prime}^{}
\Bigr]
=
 - h_{\ell\ell^\prime}^{}
 \Bigl[
  \overline{(\ell_L)^c}\, \ell_L^\prime \Delta^{++}
 \Bigr]
 -
 \sqrt{2}\, h_{\ell\ell^\prime}^{}
 \Bigl[
  \overline{(\nu_{\ell L}^{})^c}\, \ell_L^\prime \Delta^+
 \Bigr]
 +
 h_{\ell\ell^\prime}^{}
 \Bigl[
  \overline{(\nu_{\ell L}^{})^c}\, \nu_{\ell^\prime L}^{} \Delta^0
 \Bigr] ,
\end{eqnarray}
where the Yukawa coupling constants satisfy
$h_{\ell\ell^\prime}^{} = h_{\ell^\prime\ell}^{}$.
 The vev of $\Delta^0$ can generate
neutrino masses~\cite{Schechter:1980gr, Ref:HTM} as
\begin{eqnarray}
(m_\nu^{})_{\ell\ell^\prime}^{}
= 2 \langle \Delta^0 \rangle h_{\ell\ell^\prime}^{} .
\end{eqnarray}
 Hereafter,
we refer to this solo mechanism of generating neutrino masses
as the Higgs triplet model.
 Branching ratios
$\BR(\Delta^+ \to \overline{\ell_L^{}}\, \overline{\nu_L^{}})$
in the Higgs triplet model
are proportional to $(m_\nu^\dagger m_\nu)_{\ell\ell}^{}$
identically to those in the neutrinophilic two-Higgs-doublet model.
 Therefore,
the discussion in the previous subsection
is applicable also for
$\Delta^+ \to \overline{\ell_L^{}}\,\overline{\nu_L^{}}$.
 See e.g.\ Fig.~16 in Ref.~\cite{Perez:2008ha}
to compare with Fig.~\ref{Fig:H+} in this article.
 If non-leptonic decays (e.g.\ $\Delta^+ \to W^- \Delta^{++}$)
are not negligible,
a ratio of branching ratios of $e\nu$ and $\mu\nu$ modes
would be reliable.

\section{Doubly Charged Scalar}
\label{Sec:doubly}

\subsection{$\SU(2)_L$-singlet}
\label{Subsec:doubly-singlye}

 An $\SU(2)_L$-singlet scalar $s^{++}$ with $Y=2$
has the following Yukawa interaction:
\begin{eqnarray}
f_{\ell\ell^\prime}^\prime
\Bigl[ \overline{(\ell_R)^c}\, \ell_R^\prime\, s^{++} \Bigr] ,
\end{eqnarray}
where the Yukawa coupling constants satisfy
$f_{\ell\ell^\prime}^\prime = f_{\ell^\prime\ell}^\prime$.
 The scalar $s^{++}$ is introduced in e.g.\
the Zee-Babu model~\cite{Ref:ZBM}
where an $\SU(2)_L$-singlet scalar $s^+$
(see also Sections~\ref{Subsec:singly-singlet-I}
and \ref{Subsec:singly-singlet-II})
is also introduced.
 The Yukawa interaction with $s^+$ is shown in Eq.~\eqref{Eq:s-Yukawa}.

 Motivated by the Zee-Babu model,
let us take a case in which
the structure of neutrino mass matrix $(m_\nu^{})_{\ell\ell^\prime}$
is given by
\begin{eqnarray}
(m_\nu^{})_{\ell\ell^\prime}
\propto
 \left[
  f m_\ell^\diag f^\prime m_\ell^\diag f^T
 \right]_{\ell\ell^\prime} ,
\label{Eq:mnu-ZBM}
\end{eqnarray}
where $m_\ell^\diag \equiv \diag( m_e, m_\mu, m_\tau )$.
 Discussion in this subsection is based on
Refs.~\cite{Ref:ZBM-LHC-1, Ref:ZBM-LHC-2}
where the Zee-Babu model was studied.
 The lightest neutrino becomes massless ($m_1=0$ or $m_3=0$)
because of $\text{Det}(m_\nu^{}) \propto \text{Det}(f) = 0$.
 We simply assume that $m_e$ in Eq.~\eqref{Eq:mnu-ZBM} can be ignored
so that we can have some "prediction"
on the flavor structure of branching ratios
$\BR(s^{++} \to \overline{\ell_R^{}}\, \overline{\ell_R^\prime})$
which are proportional to $|f_{\ell\ell^\prime}^\prime|^2$.
 Then,
because of a large mixing angle $\theta_{23}$,
we expect
$|(m_\nu^{})_{\mu\mu}^{}|
 \simeq |(m_\nu^{})_{\mu\tau}^{}|
 \simeq |(m_\nu^{})_{\tau\tau}^{}|$
which results in
$|f_{\mu\mu}^\prime| m_\mu^2/m_\tau^2
 \simeq |f_{\mu\tau}^\prime| m_\mu/m_\tau
 \simeq |f_{\tau\tau}^\prime|$.
The branching ratios become
\begin{eqnarray}
\BR(s^{++} \to \overline{\mu_R^{}}\, \overline{\mu_R^{}})
: \BR(s^{++} \to \overline{\mu_R^{}}\, \overline{\tau_R^{}})
: \BR(s^{++} \to \overline{\tau_R^{}}\, \overline{\tau_R^{}})
\simeq 1 : 0 : 0 .
\label{Eq:BR-ZBM}
\end{eqnarray}
 If Eq.~\eqref{Eq:BR-ZBM} turns out to be consistent with measurements,
the neutrino mass matrix might be the structure in Eq.~\eqref{Eq:mnu-ZBM}
where $m_1=0$ or $m_3=0$ is predicted.

\subsection{$\SU(2)_L$-triplet}
\label{Subsec:doubly-triplet}

\begin{figure}
\begin{center}
\includegraphics[scale=0.33,angle = -90]{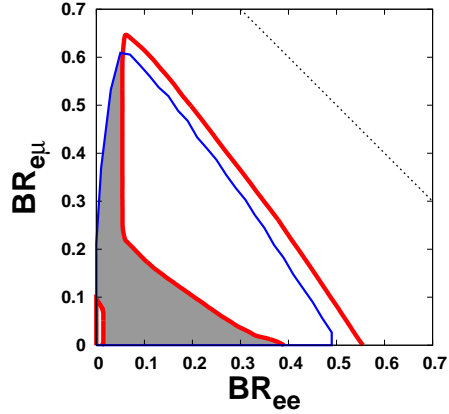}{}
\end{center}
\caption{
 Two axis are defined as
$\BR_{\ell\ell^\prime}
\equiv \BR(\Delta^{++} \to \overline{\ell_L^{}}\, \overline{\ell_L^\prime})$.
 Values of $\BR_{\ell\ell^\prime}$ in the shaded region
cannot be achieved in the model
when the lightest neutrino mass is zero.
 This figure is a simplified version of the one
in Ref.~\cite{Ref:mnu-HTM-LHC-2}.
}
\label{Fig:HTM}
\end{figure}

 A doubly charged scalar exists also in the Higgs triplet model
(see also Sec.~\ref{Subsec:singly-triplet}).
 Let us take a scenario that $\Delta^{++}$ dominantly decays
into a pair of same-signed leptons:
$\Delta^{++} \to \overline{\ell_L^{}}\, \overline{\ell_L^\prime}$.
 Then branching ratios of leptonic decays
are determined by $h_{\ell\ell^\prime}^{}$,
and the flavor structure of the branching ratios can provide
direct information on the neutrino mass matrix%
~\cite{Ref:mnu-HTM-LHC-1, Ref:mnu-HTM-LHC-2}.
 For example,
if branching ratios
$\BR(\Delta^{++} \to \overline{e_L^{}}\, \overline{e_L^{}})$
and $\BR(\Delta^{++} \to \overline{e_L^{}}\, \overline{\mu_L^{}})$
are observed in the shaded region in Fig.~\ref{Fig:HTM},
it would be excluded that
the lightest neutrino mass is zero in this model~\cite{Ref:mnu-HTM-LHC-2}.
 Note that information on the lightest neutrino mass
cannot be obtained by neutrino oscillation measurements.

\section{Summary}
\label{Sec:summary}

 We discussed relations between
the neutrino mass matrix
and the flavor structure of decays of leptophilic charged scalars.
 By assuming how a matrix of Yukawa coupling constants
for a leptophilic scalar appears in the neutrino mass matrix,
we obtained predictions on the leptonic decays of the scalar.

 If the antisymmetric matrix $f$ of Yukawa coupling constants
for an $\SU(2)_L$-singlet singly-charged scalar $s^+$
appears in the neutrino mass matrix as
$(m_\nu^{})_{\ell\ell^\prime}^{}
\simeq
 m_\tau (X_{\ell\tau} f_{\tau\ell^\prime}
 + f_{\ell\tau} X_{\tau\ell^\prime})$,
a combination of branching ratios
for $s^+ \to \overline{\ell_L^{}} \overline{\nu_L^{}}$
satisfies
$\BR_\tau^{(s)}/(\BR_e^{(s)}-\BR_\mu^{(s)}) \simeq 1$.
 For 
$\sin^2{\theta_{23}} = 1/2$ and $\sin^2{\theta_{12}} = 1/3$,
we obtained
$\sin^2{2\theta_{13}} \simeq 0.11$,
$\delta \simeq \pi$, and
$\alpha_{21}^{} \simeq \pi$.
 If the matrix $f$ appears as $m_\nu = f X_s f^T$ or $X f$,
we predicted
$\BR_e^{(s)} : \BR_\mu^{(s)} : \BR_\tau^{(s)} = 2 : 5 : 5$ for $m_1=0$
and $2 : 1 : 1$ for $m_3 = 0$.
 On the other hand,
if a leptophilic singly-charged scalar is
a member of $\SU(2)_L$-doublet or triplet,
a branching ratio for a decay into
$\overline{e_L^{}}\, \nu$ (and $\overline{\mu_L^{}}\, \nu$)
would provide information on the value of the lightest neutrino mass
and on whether $m_1 < m_3$ or not.

 The symmetric matrix $f^\prime$ of Yukawa coupling constants
for an $\SU(2)_L$-singlet doubly-charged scalar $s^{++}$
might contribute to the neutrino mass matrix as
$m_\nu^{} = f m_\ell^\diag f^\prime m_\ell^\diag f^T$.
 When we assume that a contribution of $m_e$ to $m_\nu$ is negligible
and $|(m_\nu^{})_{\mu\mu}^{}|
 \simeq |(m_\nu^{})_{\mu\tau}^{}|
 \simeq |(m_\nu^{})_{\tau\tau}^{}|$,
decays of $s^{++}$ into
$\overline{\mu_R^{}}\, \overline{\tau_R^{}}$
and $\overline{\tau_R^{}}\, \overline{\tau_R}$
become negligible in comparison with the $\mu\mu$ mode.
 For the case of $\Delta^{++}$ in an $\SU(2)_L$-triplet field,
the flavor structure of
$\Delta^{++} \to \overline{\ell_L^{}}\, \overline{\ell_L^\prime}$
directly relates to the neutrino mass matrix.
 We showed that information on the lightest neutrino mass
(and Majorana phases etc.) could be obtained
by observing the structure of the $\Delta^{++}$ decays.

 We hope that the mechanism of the neutrino mass generation is uncovered
by discovery of such leptophilic scalars
at collider experiments.
 Peaks of their signals may remind us of the Tateyama peaks in Toyama!

\begin{acknowledgments}
 I thank A.G.~Akeroyd, M.~Aoki, S.~Kanemura, and T.~Nabeshima
for collaboration on which this article is partially based.
 I also thank members of the Higgs working group
and participants in this workshop
for valuable discussions.
 This work was supported in part by the Grant-in-Aid
for Young Scientists~(B) No.~23740210.
\end{acknowledgments}

\bigskip 

\end{document}